\documentclass[fleqn,twoside]{article}
\usepackage{espcrc2}
\usepackage{times}

\usepackage{graphicx}
\usepackage[figuresright]{rotating}

\newcommand{\AmS}{{\protect\the\textfont2
  A\kern-.1667em\lower.5ex\hbox{M}\kern-.125emS}}

\title{Are magnetic monopoles hadrons?}
\author{ 
Michael Creutz 
\address 
{Physics Department, Brookhaven
National Laboratory,
\\ Upton, NY 11973, USA
} 
\thanks
{This manuscript
has been authored under contract number DE-AC02-98CH10886 with the
U.S.~Department of Energy.  Accordingly, the U.S. Government retains a
non-exclusive, royalty-free license to publish or reproduce the
published form of this contribution, or allow others to do so, for
U.S.~Government purposes.}
}
       
\begin{document}

\begin{abstract}
The charges of magnetic monopoles are constrained to a multiple of
$2\pi$ times the inverse of the elementary unit electric charge.  In
the standard model, quarks have fractional charge, raising the
question of whether the basic magnetic monople unit is a multiple of
$2 \pi/e$ or three times that.  A simple lattice construction shows
how a magnetic monopole of the lower strength is possible if it
interacts with gluonic fields as well.  Such a monopole is thus a
hadron.  This is consistent with the construction of magnetic
monopoles in grand unified theories.
\vspace{1pc}
\end{abstract}

\maketitle

\input epsf
\input colordvi  
\def \li {\par\hskip .1in \Green{$\bullet$}\hskip .1 in \textBlue}

\def \hi {\medskip \textBlack}
\def \topic #1{\par\textRed\centerline {#1}}

\long \def \blockcomment #1\endcomment{}


\def \li {\par\hskip .1in \Green{$\bullet$}\hskip .1 in \textBlue}

\def \hi {\medskip \textBlack}
\def \topic #1{\par\textRed\centerline {#1}}

\def\fitframe #1#2#3{\vbox{\hrule height#1pt
 \hbox{\vrule width#1pt\kern #2pt
 \vbox{\kern #2pt\hbox{#3}\kern #2pt}
 \kern #2pt\vrule width#1pt}
 \hrule height0pt depth#1pt}}

\long \def \blockcomment #1\endcomment{}

\def\slashchar#1{\setbox0=\hbox{$#1$}           
   \dimen0=\wd0                                 
   \setbox1=\hbox{/} \dimen1=\wd1               
   \ifdim\dimen0>\dimen1                        
      \rlap{\hbox to \dimen0{\hfil/\hfil}}      
      #1                                        
   \else                                        
      \rlap{\hbox to \dimen1{\hfil$#1$\hfil}}   
      /                                         
   \fi}                                         %


Some time ago G. 't Hooft \cite{'tHooft:1975zi} discussed an amusing
puzzle involving the quantization of magnetic monopole charges.  As is
well known from Dirac \cite{dirac}, quantum mechanical consistency for
a hypothetical monopole requires its charge to be quantized in units
of ${2\pi\over e}$, where $e$ is the smallest non-zero charge of any
existing particle.  The puzzle arises on considering quarks, which
have charges that are third integer fractions of the charges on free
particles.  Is the quantization unit ${2\pi\over e}$ or thrice that?

The resolution \cite{'tHooft:1975zi} is entwined with the phenomenon
of confinement and the strong force.  Indeed, the minimum charge is
${2\pi\over e}$, but to have this value, a magnetic monople must also
interact with gluon fields.  Here I return to this old result and
discuss it in simple lattice gauge language.

I begin the discussion with a description of how a static magnetic
monopole is formulated in pure U(1) lattice gauge theory.  Since it is
static, I ignore the time direction, along which all links and
plaquettes are unmodified.  The following structure is repeated on
each time slice.

To start, use a ``Dirac string'' to bring in from infinity the net
flux emerging from the monopole.  For the gauge fields, there is a
phase $U_l$ associated with each link on our lattice.  The gauge
action is constructed by multiplying links around plaquettes
$U_p=\prod_{l \in p}U_l$.  Ordinarily the action just adds the real
parts of these plaquette variables together.  But with the Dirac
string present, we insert an additional phase on plaquettes pierced by
the string and take
\begin{equation}
S=\sum_p {\rm Re}(U_p e^{i\phi_p}).
\end{equation}
Here $\phi_p$ is the external flux brought through the plaquette and
vanishes except along the string ending at the monople location.

\begin{figure}[htb]
\centering
\includegraphics[width=1.7in]{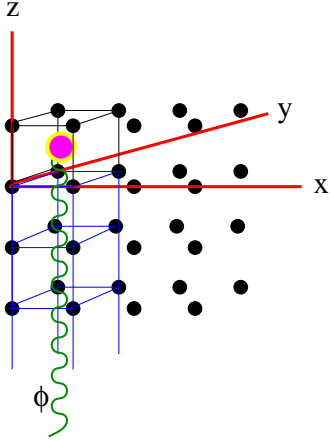}
\caption{\label{fig:monopole} The construction of a static monopole at
the origin.  The monopole flux is brought in along a Dirac string
parallel to the negative $z$ axis.}
\end{figure}

So, to have a monopole centered in a cube at the origin, one can run
the string down the negative $z$ axis.  Then  $\phi_p=0$ except on 
$xy$ plaquettes at $x=y=0$, $z\le 0$.  For those plaquettes, $\phi_p$ is
a constant and represents the monopole strength.  This is illustrated
in Fig.~(\ref{fig:monopole}).

\begin{figure}[htb]
\centering
\includegraphics[width=1.3in]{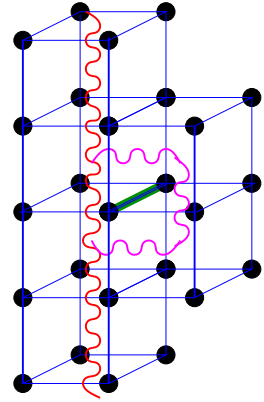}
\caption{\label{fig:kink} The path taken by a Dirac string is
unphysical since it can be rerouted by simple changes of variables on
the lattice links.}
\end{figure}

The route taken by the Dirac string can be changed by simple changes
of variables.   For example, if we absorb the factor of $e^{i\phi_p}$
on some plaquette into the measure for one of the links on its side,
we reroute the Dirac string through two neighboring cubes, as
illustrated in Fig.~(\ref{fig:kink}).

Note that if the monople has strength $\phi=2\pi$, the action is
equivalent to the pure gauge theory without the monopole present.  As
interpreted by Degrand and Toussaint \cite{DeGrand:1980eq}, the
compact theory has intrinsic monopoles of strength $2\pi$ which
completely screen the external monopole.

So far I have been discussing the theory without dynamical charges.
Now include an electrically charged fermion of intrinsic charge {$e$}.
When this fermion hops between neightbors along link $l$, its wave
function picks up a phase $\left(U_l\right)^e$.  Unlike the pure gauge
situation, the Dirac string with arbitrary strength becomes
observable.  When the fermion hops around it, it picks up an extra
phase $e^{ie\phi}$.  Only if $e\phi=2\pi n$ is the string
unobservable.  This is the Dirac quantization condition.

\begin{figure}[htb]
\centering
\includegraphics[width=1.7in]{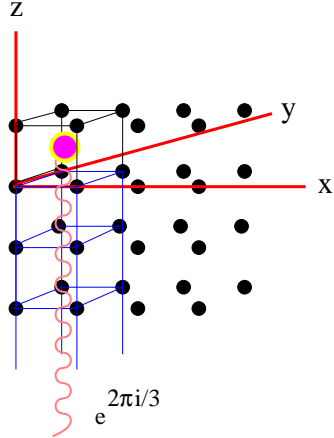}
\caption{\label{fig:z3} In analogy with the $U(1)$ monopole, a $Z_3$
monopole can be constructed for the $SU(3)$ gauge theory of the strong interactions.}
\end{figure}

Now I turn to discuss another type of string, a $Z_3$ string in the
quark confining dynamics of the strong interactions.  The complex
phase $e^{2\pi i/3}$ is an element of $SU(3)$.  It commutes with all
$SU(3)$ elements, and generates the center of the group.  We can use
this element to generate strong ``monopoles'' in the pure $SU(3)$
gauge theory.  This is done in direct analogy with the above
construction of $U(1)$ monoples; a flux of strength $e^{2\pi i/3}$ is
brought in along a new ``Dirac string'' by inserting this $Z_3$ factor
into plaquettes pierced by the string.  This is illustrated in
Fig.(\ref{fig:z3}).

In the pure glue theory, the path of this string can also be changed
by simple changes of variables.  Thus the path is, as in the $U(1)$
case, unphysical.  However, when quarks are introduced into the
theory, the Dirac string becomes observable due to the extra $Z_3$
factor they encounter on circumnavigating the string.

Next consider the combined $U(1)\times SU(3)$ model.  We have on each
link $l$ both a phase factor $U_l\in U(1)$ and a a strong factor
$V_l\in SU(3)$.  The $U(1)$ and $SU(3)$ fields do not directly
interact with each other, but they are coupled via the quarks, which
interact with both.

\begin{figure}[htb]
\centering
\includegraphics[width=1.7in]{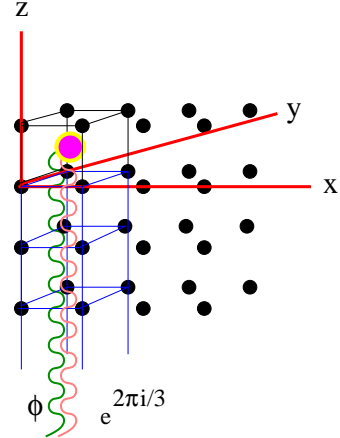}
\caption{\label{fig:superpose} Superposing $U(1)$ and $SU(3)$ strings
allows a construction invisible to both electrons and quarks.}
\end{figure}

In the coupled theory, consider superposing the $U(1)$ and the $SU(3)$
strings, as sketched in Fig.~(\ref{fig:superpose}).  Suppose the
$U(1)$ strength is $\phi=2\pi/e$ so as to make the string invisible to
electrons.  Of course, the electrons are neutral with respect to
gluons and do not interact with the $SU(3)$ string.

Quarks, interacting both with gluons and photons, do interact with
both strings.  For example, the down quark gets factor of $e^{-2\pi
i/3}$ from the $U(1)$ string.  But it also gets a factor of $e^{2\pi
i/3}$ from the $Z_3$ string.  We have a remarkable phase cancellation
between these factors, and the combined string is not observable.
Note that this cancellation locks in the quark charges, {\it i.e.} it
only works if $e_d=-e/3$ modulo full units of electric charge.  This
connection with the quark electromagnetic charge is automatic in grand
unified theories \cite{Georgi:1972cj,'tHooft:1974qc}.

\begin{figure}[htb]
\centering
\includegraphics[width=2in]{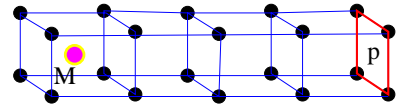}
\caption{\label{fig:correlation} 
A plaquette near a monopole has its expectation value modified.  In
the strong coupling limit, this modification appears in a diagram
similar to that used to calculate the glueball mass.}
\end{figure}

Since this minimally charged monopole involves the $SU(3)$ fields, it
is strongly interacting.  In particular, it will disturb the
surrounding gluonic fields.  For example, the plaquette expectation
value will be modified in the vicinity of the monopole.  This can be
easily seen in the strong coupling expansion, where this correlation
appears in a diagram involving tiling a tube contaning the given
plaquette and also surrounding the monopole.  This diagram is shown in
Fig.(\ref{fig:correlation}), and is quite similar to the strong
coupling diagram for the glueball mass $M_g$.  The leading behavior
takes the form
\begin{equation}
\langle U_p\rangle_M - \langle U_p \rangle_0 
\sim \beta^{4L} (e^{2\pi i/3}-1) \sim e^{-M_gL}.
\end{equation}
To this order the screening length for the hadronic fields about a
monopole is given by the glueball mass.  At higher orders diagrams
involving pions will give the longest range behavior.

To summarize, magnetic monopole charges are quantized in units of
{$2\pi/e$}, not $3\times 2\pi/e$.  But to have this value, a minimally
charged monopole must interact strongly.  To leading order, the
magnetic glue screening length is controlled by glueball mass.
Finally, this scheme works only if quark charges are fixed at their
usual values, as in grand unified theories.  While this hints at
unification, it does not require such.

\end{document}